\documentclass[twocolumn,showpacs,amsmath,amssymb,prl,nofootinbib]{revtex4}

\usepackage[all,2cell]{xy}
\usepackage{color}
\usepackage[latin1]{inputenc}
\usepackage{epsfig}
\usepackage{transparent}
\usepackage{nicefrac}
\usepackage{amsmath}
\usepackage{amssymb}
\usepackage{chemarrow}
\usepackage{bm}
\usepackage{graphicx}
\usepackage{dcolumn}
\usepackage{hyperref,color}

\newcommand{\be}{\begin{equation}}
\newcommand{\ee}{\end{equation}}
\newcommand{\bea}{\begin{eqnarray}}
\newcommand{\ba}{\begin{array}}
\newcommand{\eea}{\end{eqnarray}}
\newcommand{\bes}{\begin{subequations}\bea}
\newcommand{\ees}{\eea\end{subequations}}
\newcommand{\ea}{\end{array}}

\newcommand{\bra}[1]{\langle \, #1 \,|}
\newcommand{\ket}[1]{|\,#1\,\rangle}
\newcommand{\braket}[2]{\langle \,#1\,|\,#2\,\rangle}

\newcommand{\tb}[1]{\textcolor{black}{#1}}

\newcommand{\Y}{Y}
\newcommand{\X}{X}
\newcommand{\C}{C}
\newcommand{\matr}{A}

\begin{document}

\title{Conservation laws and symmetries in stochastic thermodynamics}

\author{Matteo Polettini} 
\affiliation{Complex Systems and Statistical Mechanics, Physics and Materials Science Research Unit, University of Luxembourg, Campus
Limpertsberg, 162a avenue de la Fa\"iencerie, L-1511 Luxembourg (G. D. Luxembourg)} 
 
\author{Gregory Bulnes-Cuetara } 
\affiliation{Complex Systems and Statistical Mechanics, Physics and Materials Science Research Unit, University of Luxembourg, Campus
Limpertsberg, 162a avenue de la Fa\"iencerie, L-1511 Luxembourg (G. D. Luxembourg)} 

\author{Massimiliano Esposito} 
\affiliation{Complex Systems and Statistical Mechanics, Physics and Materials Science Research Unit, University of Luxembourg, Campus
Limpertsberg, 162a avenue de la Fa\"iencerie, L-1511 Luxembourg (G. D. Luxembourg)} 

\date{\today}

\begin{abstract}
Phenomenological nonequilibrium thermodynamics describes how fluxes of conserved quantities such as matter, energy and charge flow from outer reservoirs across a system, and how they irreversibly degrade from one form to another. Stochastic thermodynamics is formulated in terms of probability fluxes circulating in the system's configuration space. The consistency of the two frameworks is granted by the condition of local detailed balance, which specifies the amount of physical quantities exchanged with the reservoirs during single transitions between configurations. We demonstrate that the topology of the configuration space crucially determines the number of independent thermodynamic affinities (forces) that the reservoirs generate across the system, and provide a general algorithm that produces the fundamental affinities and their conjugate currents contributing to the total dissipation, based on the interplay between macroscopic conservations laws for the currents and microscopic symmetries of the affinities.
\end{abstract} 

\pacs{05.70.Ln,02.50.Ga}


\maketitle

Thermodynamics is the science of nonequilibrium processes occurring in open systems that interact with an environment.  Today, a dramatic evolution is reshaping it, from a patchwork of general principles and applied laws --- and a riddle for students, from a pedagogical perspective --- to a systematic and comprehensive theory called Stochastic Thermodynamics (ST), where all propositions are well-founded on the mathematics of Markov processes \cite{seifertrep,broeck,jarzynski,espoCG}, with a span of applications ranging from molecular motors \cite{bernetal} to photovoltaic cells \cite{nitzan} and beyond. Still some conceptual leaps need to be filled before this program can be deemed complete. According to classic formulations \cite{degroot,prigogine}, phenomenological thermodynamics is a discourse about fluxes of energy, matter, charge etc., their {\it conservation}, and their {\it degree of degradation}, quantified by the entropy production rate (EPR).  The conceptual pathway to nonequilibrium processes starts from an ideally isolated universe, where Noether's theorem states that conservation laws follow from symmetries of the dynamics. Nonequilibrium behavior ensues when one can separate the universe into a system and its environment, which is eventually structured into several competing ideal reservoirs that always remain at equilibrium. The system's effective dynamics becomes dissipative, but, as we will argue, its features still bear the signature of the conservation laws across the system/environment interface.  

More specifically, let us consider a ``black box'' scenario where we only know that a system is in a nonequilibrium steady state generated by $R$ reservoirs denoted $r$ and described by $n_\Y$ different {\it affinities} $f_y$ (intensive thermodynamic variables such as inverse temperatures $\beta_r$ or chemical potentials $-\beta_r \mu_r$), which we list in a vector $\ket{f_\Y}$. From phenomenological thermodynamics, the EPR of this setup quantifies the entropy changes in the reservoirs caused by the $n_\Y$ physical {\it currents} $j_y$ of conserved quantity $y$ conjugated to the affinities (currents of extensive quantities such as energy $-\dot{\epsilon}_r$ or matter $-\dot{n}_r$). By convention these currents enter the reservoirs and are listed as a vector $\ket{j_\Y}$. According to the fundamental relation, each conjugate pair $f_y j_y$ is a contribution to the entropy change of a reservoir, meaning that their sum over all $n_\Y$ pairs is the physical EPR, denoted $\sigma_\Y = \braket{f_\Y}{j_\Y} \geq 0$. At this level of description, beside global conservation laws for like currents, (e.g. $\sum_r \dot{\epsilon}_r=0$ or $\sum_r \dot{n}_r=0$), no other argument can be used to further simplify the EPR. It will thus display a number of currents and affinities equal to at most $n_\Y$ minus the number of conserved quantities.

ST instead allows to ``enter the box''. It describes the system degrees of freedom as nodes of a network and its dynamics as a Markov process driven by transition probability rates, associated to network edges. Pairs of nodes may be connected by multiple edges when different mechanisms (i.e. sets of reservoirs) trigger the transition. At that level, the only conservation law is that of probability and the dynamics is characterized by the statistical EPR which additively measures the breakage of detailed balance in each edge of the network, and will be denoted $\sigma_\X = \braket{f_\X}{j_\X} \geq 0$, where $\bra{f_\X}$ and $\ket{j_\X}$ are respectively vectors of edge affinities and currents.
The crucial ingredient connecting the statistical and physical levels of description in ST is the {\it local detailed balance} (LDB) condition \cite{seifertrep, ldbmass, ldbmaes}. It relates rates to the physical quantities exchanged with the reservoirs, in such a way that $\sigma_\X$ will eventually be solely expressed in terms of physical currents. However, identifying the fundamental currents and affinities that contribute to the EPR can only be done ``by hand'' in very simple systems and no systematic procedure exists to address this crucial question in more complex ones. An important step forward was made by Schnakenberg \cite{schnak}, who showed that the statistical EPR can be expressed as a sum over the number of fundamental cycles $n_\X$ of the network, of products of the cycle affinities $F_\X$ and currents $J_\X$, i.e. $\sigma_\X = \braket{F_\X}{J_\X}$. Alas, the number of configuration cycles typically grows exponentially large with the network size, and most importantly many of these cycle affinities are not independent of each other.  It is then crucial to overcome this major limitation in Schnakenberg's analysis.

In this Letter, by adapting the formalism of closed and open chemical networks proposed in Ref.\,\cite{polettiniCN}, we provide a general and systematic procedure for doing so. Beside its conceptual aspects, bringing light on the trade-off between symmetries and conservation laws --- a mechanism that is somewhat reminiscent of the Noether theorem, our procedure paves the the way to applications of ST to systems with arbitrarily large and complex networks. 
 
	\begin{figure*}[htb]
  	\centering
 \includegraphics[width=\textwidth]{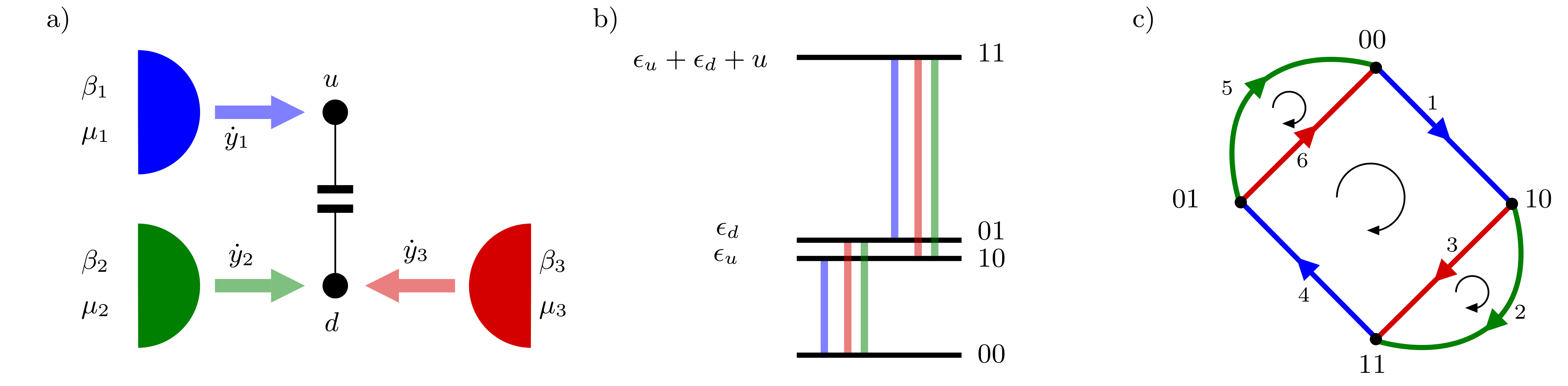}
\caption{ \label{quantumdots} a) Physical representation of two quantum dots capacitively coupled and in contact with three reservoirs, $\dot{y}_r = (\dot{\epsilon}_r,\dot{n}_i)$; b) Many-body energy levels and their single-level occupation numbers; c) Network representation of the configuration space where the Markov process occurs, with an arbitrary orientation assigned to each edge.}
	\end{figure*}

We start by illustrating our argument using a simple setup. The system consists of two single-level quantum dots, coupled among themselves by an effective capacitance $C$, and exchanging energy and particles with three electronic reservoirs $r=1,2,3$, each at different inverse temperatures $\beta_r$ and chemical potentials $\mu_r$, according to Fig.\ref{quantumdots}a), for a total of $n_\Y=6$ affinities. The system's dynamics can be described as a continuous-time Markov jump process with rates describing transitions between states $x$ in the system space of configurations $X = \{00,01,10,11\}$ which correspond to the absorption/emission of energy and particles from the reservoirs, as depicted in Fig.\,\ref{quantumdots}b) \cite{strasberg,bulnes,sanchez}. Each possible transition mechanism belongs to the space of the oriented edges of a network (or graph) that has $X$ as its vertices (see Fig.\,\ref{quantumdots}c). The rates satisfy local detailed balance and a simple stochastic thermodynamics ensues.
When regarding the system as a ``black box'' that only serves to process energy fluxes $\dot{\epsilon}_r$ and particle fluxes $\dot{n}_r$ between the reservoirs, physical EPR reads $\sigma_\Y = - \sum_{r=1}^3 \beta_r (\dot{\epsilon}_r - \mu_r \dot{n}_r) = \braket{f_\Y}{j_\Y}$, where the current and affinity vectors are respectively $\ket{j_\Y} = (-\dot{\epsilon}_r,-\dot{n}_r)$ and $\bra{f_\Y} = (\beta_r,-\mu_r\beta_r)$. One can now make use of conservation of energy $\dot{\epsilon}_1 + \dot{\epsilon}_2 + \dot{\epsilon}_3 = 0$ and of particles $\dot{n}_1 + \dot{n}_2 + \dot{n}_3 =0$ to simplify the EPR as 
$\sigma_\Y = \sum_{r=1}^2  \big[ \left(\beta_3 - \beta_r \right) \dot{\epsilon}_r + (\beta_r\mu_r - \beta_3\mu_3) \dot{n}_r \big]$
and thus reduce the number of affinities from $n_\Y=6$ to $4$. The ``black box'' perspective does not allow to go further. However, when considering the internal structure of the system, one notices that electrons cannot cross the condensator. Hence there is one additional conservation law $\dot{n}_1 = 0$ which allows to further simplify the EPR by reducing one more affinity
\begin{multline}
\sigma_\Y =
\sum_{r=1}^2  \left(\beta_3 - \beta_r \right) \dot{\epsilon}_r + (\beta_2\mu_2 - \beta_3\mu_3) \dot{n}_2 = \braket{F_\Y}{J_\Y} \label{eq:2l2}.
\end{multline}
We thus learned that out of the $n_\Y=6$ affinities describing the reservoirs, only $3$ fundamental affinities $\bra{F_\Y}$ and their conjugated currents $\bra{J_\Y}$ ultimately control the physical EPR, due to $2$ global and $1$ system-specific conservation laws. While identifying the latter was easy in this simple setup, doing so in more complex systems is not and requires a systematic procedure.

We now proceed with the general theory. We consider a system dynamics described by the master equation $\partial_t p_x = \sum_{\nu ,x'} \left( w^{\nu}_{xx'} p_{x'} - w^{\nu}_{x'x} p_x \right)$, where $\nu$ distinguishes between different transitions that connect two states. The network EPR $\sigma_\X$ is defined as \cite{schnak}
\bea
\sigma_\X 
= \frac{1}{2} \sum_{x,x',\nu} \stackrel{j^{\X}_{xx',\nu}}{\overbrace{\hspace{-1.2cm} \phantom{\ln \frac{w^{\nu}_{xx'}}{w^{\nu}_{x'x}}} \Big( w^{\nu}_{xx'} p_{x'} - w^{\nu}_{x'x} p_x \Big) }} \; \stackrel{f^{\X}_{xx',\nu}}{\overbrace{\ln \frac{w^{\nu}_{xx'} p_{x'}}{w^{\nu}_{x'x} p_{x}}}} \; \geq 0
\eea
where the overbraces respectively define the probabilistic currents and their conjugate affinities. Letting $e=(xx',\nu)_{x < x'}$ label the edges of the graph, the {\it incidence matrix} $\matr^\X$ of the network has entries
\bea
\matr^\X_{x,e} = \left\{
\ba{ll}
+1 & \mbox{ if } \stackrel{e}{\longrightarrow} x  \\
- 1 & \mbox{ if } \stackrel{e}{\longleftarrow} x \\
0 & \mbox{ otherwise } 
\ea
\right. .
\eea
The master equation can then be cast in the form of a continuity equation $\partial_t \ket{p} = \matr^\X \ket{j_\X}$. 
We will focus on steady states, where Kirchhoff's Current Law holds $\matr^\X \ket{j_\X} = 0$, implying that  $\ket{j_\X}$ lives in the null space of the incidence matrix, which is known to be spanned by $n_\X$ independent cycles of the graph. Schnakenberg \cite{schnak} described a procedure (that we call {\bf routine 1}, see Refs. \cite{schnak,gaspard,polettiniprojectors} for details) to find a basis of cycle vectors. The steady network currents can be expressed as $\ket{j_\X} = \matr^\C \ket{J_\X}$, where $\matr^\C$ is a full-rank matrix of independent null vectors of $\matr^\X$, viz. $\matr^\X \matr^\C =  0$, and $\ket{J_\X}$ is a vector of coefficients with the meaning of independent cycle currents. Notice that there is a certain degree of freedom in the choice of $\matr^\C$. 
Defining the cycle affinities $\bra{F_\X} = \bra{f_\X} \matr^\C$, we obtain the well-known decomposition of the network EPR
\bea
\sigma_\X = \braket{f_\X}{j_\X} = \braket{F_\X} {J_\X}, \label{eq:EPR}
\eea
where it is important to notice that the  affinity of a cycle $\gamma$ only depends on the rates, $F_\X(\gamma) = \ln \prod_{e \in \gamma} \frac{w_{e}}{w_{-e}}$. Notice that both at the network and at the physical level we resort to uppercase symbols $J,F$ when we take into account the respective conservation laws (of probability, or of physical quantities).


The passage from statistical to physical thermodynamics is based on the identification of physical currents as linear combinations of network currents:
\bea
\ket{j_\Y} = \matr^\Y \ket{j_\X} 
\label{eq:defppj}
\eea
where $\matr^\Y_{y,e} = y_x - y_{x'}$ is the inflow of extensive quantity $y$ as the system performs a transition $e = (xx',\nu)$. Steady-state
{\it thermodynamic consistency} requires that the network EPR $\sigma_\X$ coincides with the physical one $\sigma_\Y$, namely there exists a vector of phenomenological thermodynamic affinities $\bra{f_\Y}$ such that
\bea
\sigma_{\Y} :=\braket{f_\Y}{j_\Y} = \braket{f_\X}{j_\X}= \braket{F_\X} {J_\X}. \label{eq:ThermoCons}
\eea
Steady-state thermodynamic consistency is granted by the following condition of LDB,
\bea
\bra{f_{\X}}=\bra{f_{\Y}} \matr^{\Y} + \bra{\Delta\phi}  \label{eq:thfor}
\eea
on the assumption that $ \bra{\Delta \phi}  A^\C = 0$. This formula then relates the edge affinities to the reservoir entropy changes caused by the transition along that edge, up to an increase in a state function $\phi$ that measures the system entropic changes. We discuss specific examples below.

Introducing the $n_\Y \times n_\X$ matrix 
\bea
M := \matr^\Y \matr^\C
\eea
 we find
\bea
\ket{j_\Y} = M \ket{J_\X}  \ \ \;, \ \ \bra{F_\X} = \bra{f_\Y} M \label{eq:pf}.
\eea
This shows that $M$ is the crucial object to understand the mapping between physical and cycle thermodynamics, as it mixes ``black box'' information and topology. Importantly, the passage from physical to cycle EPR in Eq. (\ref{eq:ThermoCons}) comes with a balance of conservation laws and symmetries. On the one hand, letting $\bra{w}$ be any of the $\lambda_\Y$ independent left-null vectors of $M$, then $\braket{w}{j_\Y} = 0$
which expresses the conservation of physical currents across the system. 
On the other hand, for each of the $\lambda_\X$ right-null vectors $\ket{v}$ of $M$ we have $\braket{F_\X}{v} = 0$, which expresses symmetries of the cycle affinities.
Notice that the inverse problem of reconstructing $\bra{f_\Y}$ from the cycle affinities $\bra{F_\X}$, for given $\phi$, is not uniquely determined, and as a consequence one can further compress the expression for the EPR. Since the rank of $M$ is
\bea
\alpha := n_\Y - \lambda_\Y = n_\X - \lambda_\X , \label{eq:alpha}
\eea
the EPR can be expressed as 
\bea
\sigma = \braket{F_\Y}{J_\Y}
\eea
in terms of a reduced number $\alpha$ of fundamental physical currents $\ket{J_\Y}$ and affinities $\bra{F_\Y}$. 
The latter two equations summarize our main findings: 
The EPR of a system at steady state between different reservoirs with $n_\Y$ affinities only displays $\alpha = n_\Y - \lambda_\Y$ fundamental affinities and currents because the internal structure of the system enforces $\lambda_\Y$ (i.e. the dimension of the cokernel of $M$) conservation laws. We also improve the Schnakenberg construction since the number of symmetries $\lambda_\X$ (i.e. the dimension of the kernel of $M$) determines how many of the $n_\X$ cycle currents and cycle affinities are redundant.
The balance Eq.\,(\ref{eq:alpha}) shows that in a given network, for fixed $n_\X$ and $n_\Y$, when varying the reservoir affinities and thus the rates, the eventual appearance of an additional conservation law necessarily comes with the simultaneous appearance of one further symmetry of the affinities, in a mechanism that is reminiscent of Noether's theorem in classical mechanics (see Ref. \cite{baez} for a different formulation of a Markovian Noether-type theorem for the probability).

We now provide a systematic procedure, {\bf routine 2}, to produce these fundamental affinities and currents. It can be seen as the analog of {\bf routine 1} at the physical level: Define $W$ as the matrix of independent left-null vectors of $M$. Notice that $W \ket{j_\Y} = 0$ implies that $\ket{j_\Y} = \tilde{M} \ket{J_\Y}$, where $\tilde{M}$ is a matrix of independent right-null vectors of $W$ \tb{(e.g. obtained by removing $\lambda_Y$ columns from $M$). To find the fundamental currents we then just need to invert this relation using the Moore-Penrose pseudoinverse, $\ket{J_\Y} = \tilde{M}^{+} \ket{j_\Y} = \tilde{M}^{+}  M \ket{J_\X}$ \cite{moore}. Similarly, the fundamental affinities can be found by solving the linear equations $\bra{F_\Y} \tilde{M}^+ M = \bra{F_\X}$ on the subspace $\bra{F_{\X}} V = 0$, where $V$ is the matrix of right-null vectors of $M$. A vector space analysis shows that this routine has a unique solution (found for example by removing $\lambda_\X$ linear equations corresponding to non-independent rows of $\tilde{M}^+ M$)}. 
However, like for $\matr^\C$, there is freedom in the choice of $\tilde{M}$ and the choice of a preferred basis of null vectors must be based on the specifics of the system at hand.


Let us resume our results by the following algorithm, which for a given model, finds conservation laws and symmetries and provides an expression for the fundamental currents and affinities: 
{\it (i)} Input master equation with rates satisfying LDB; 
{\it (ii)} Find the incidence matrix $\matr^\X$; 
{\it (iii)} Find $\matr^\C$ using {\bf routine 1}; 
{\it (iv)} Calculate cycle affinities $\bra{F_X}$ and currents $\ket{J_\X}$; 
{\it (v)} Input $\matr^\Y$ compatibly with LDB; 
{\it (vi)} Compute $M=\matr^\Y \matr^\C$; 
{\it (vii)} Find symmetries as right-null vectors of $M$; 
{\it (viii)} Find conservation laws as left-null vectors of $M$; 
{\it (ix)} Use {\bf routine 2} to find the fundamental affinities $\bra{F_Y}$ and currents $\ket{J_\Y}$.


Let us now discuss some examples of LDB, as can be found e.g. in Refs. \cite{espoCG, espLDB1, espLDB2, espLDB3, artur}. When the system transitions are caused by exchanges of energy and particles $y = (\epsilon_1,\ldots,\epsilon_R,n_1,\ldots,n_R)$, one at the time, with $R$ grand-canonical reservoirs with physical affinities $\bra{f_\Y} = (\beta^1,\ldots,\beta^R,-\beta^1 \mu^1,\ldots,-\beta^R \mu^R)$, and possibly internal entropy states e.g. due to coarse graining \cite{espoCG}, the LDB condition reads \cite{ldbmass}
\bea
\ln \frac{w^r_{xx'}}{w^r_{x'x}} = \beta^r (\epsilon_{x'}-\epsilon_{x}) - \beta^r \mu^r (n_{x'}-n_{x}) + s_{x'} - s_x. \label{eq:thfor}
\eea
Energy and matter currents are given respectively by $j^\Y_{\epsilon_r} = \sum_{x,x'} (\epsilon_x - \epsilon_{x'}) j^\X_{xx',r}$, $j^\Y_{n_r} = \sum_{x,x'} (n_x - n_{x'}) j^\X_{xx',r}$. This setup can be easily shown to satisfy the above framework, with matrix $\matr^\Y$ given by 
\bea
\matr^\Y_{y,e} = \left\{ \ba{lll} \epsilon_{x'} - \epsilon_x, & \mathrm{if}\; e = x\stackrel{r}{\gets} x', & y = \epsilon_r \\ 
n_{x'} - n_x, & \mathrm{if}\; e = x\stackrel{r}{\gets} x', & y = n_r \\
0 & \mathrm{otherwise}
\ea \right.
\eea
and the potential $\phi_x$ accounting for the internal entropy $s_x$ and for the self-information $-\log p_x$. It follows from the fact that $\matr^\Y$ has a block structure (energy/particle) and that it is defined only in terms of energy differences and of particle number differences, that the maximum value of $\alpha$ is $2(R-1)$. However, additional symmetries following from the network properties might further reduce this number. They can neither be deduced from the single edge level where the LDB is expressed, nor from a global black-box perspective.  Another example is the mass-action kinetics in stochastic chemical dynamics, cf. Ref.\,\cite{artur}. 

We conclude by going back to our model system in the light of the full theory. Details are deferred to \cite{SuppMat}. A Schnakenberg analysis reveals $n_\X = 3$ cycle affinities $F_\X^1 = (\epsilon_d + u) (\beta_3 - \beta_2) + \beta_2 \mu_2 - \beta_3 \mu_3$, $F_\X^2 = \epsilon_d (\beta_2 - \beta_3) + \beta_3\mu_3 - \beta_2 \mu_2$ and $F_\X^3 = (\beta_1 -\beta_3) u$, corresponding to the three cycles depicted in Fig.\,(\ref{quantumdots}). 
The matrix $M$ reads
\bea
M = \left(\ba{ccc}
0 & 0 & u \\
-\epsilon_d - u  & \epsilon_d & 0 \\ 
\epsilon_d + u & -\epsilon_d & -u \\ 
0 & 0 & 0 \\
-1 & 1 & 0 \\
1 & -1 & 0
 \ea
\right). \label{eq:2}
\eea
Since it is full-rank, there is no symmetry of the affinity $\lambda_\X = 0$, which implies that thermodynamic consistency is granted, and that there are $\alpha = 3$ fundamental affinities and currents, and $\lambda_\Y = 3$ conservation laws corresponding to left-null vectors of $M$, namely 
\bea
W = \left(\ba{cccccc} 1 & 1 & 1 & 0 & 0 & 0 \\ 
0 & 0 & 0 & 1 & 0 & 0 \\
0 & 0 & 0 & 0 & 1 & 1 \ea
\right),
\eea
whose rows correspond respectively to total energy conservation, conservation of the number of particles in the upper quantum dot, and conservation of the number of particles in the lower quantum dot.
More interesting is the situation when we set all $\beta_r \mu_r$ identical and thus $n_\Y=3$. The physical framework then reduces to the fluxes of energy only. Matrix $M$ is given by the upper half-block in Eq.\,(\ref{eq:2}). Then, there is one conservation law $W = (1,1,1)$, one symmetry of the affinities $V^T = (\epsilon_d,\epsilon_d + u,0)$, and thus $\alpha=2$ fundamental affinities and currents. The fact that $\bra{F_\X} V = 0$ again confirms that LDB grants thermodynamic consistency. 

While this simple example could be worked out without the aid of the machinery outlined above, first-sight resolutions are impossible when the network becomes large. The utility of our approach is illustrated in \cite{SuppMat} with application to a randomized large grid.
 
\paragraph*{Conclusions.} 

We provided a systematic procedure to identify the fundamental set of currents and affinities governing the entropy production of a system in contact with multiple reservoirs, thus expanding the realm of application of stochastic thermodynamics to larger and more complex systems. 
Our theory revealed the fundamental role of the network topology on the thermodynamic description. 
Our presentation focused on ensemble averaged EPR, but our results can be directly transferred to fluctuating ST and to fluctuation theorems at the large deviation level \cite{ls,gaspard,polespo}. 

Finally, as a perspective, let's consider the linear regime. Schnakenberg computed the response matrix $L$ for the cycle observables, showing that it is symmetric and nondegenerate. In our setup, the physical linear response relation reads $\ket{f_\Y} = M^TLM \ket{j_\Y}$ as can be immediately deduced from Eq.\,(\ref{eq:pf}) and \cite[Eq.\,(10.18)]{schnak}. Matrix $M^TLM $ is symmetric, hence Onsager symmetry is granted at the macroscopic level in the presence of LDB. We notice that the existence of conservation laws is crucial for optimizing the efficiency of machines \cite{casati}. Interestingly, the above matrix becomes degenerate in presence of conservation laws; degeneracy is precisely the condition required to reach the so-called tight-coupling condition that optimizes the efficiency of machines \cite{prost}. Therefore our analysis might have interesting consequences in the study of efficiency enhancement.

\paragraph*{Aknowledgments.} 

This research was supported by the National Research Fund Luxembourg in the frame of project FNR/A11/02 and of Postdoc Grant 5856127 
and by the European Research Council (project 681456). Discussion with A. Lazarescu and with R. Rao was highly beneficial.

\end{document}